\documentclass[showpacs,amsmath,amssymb,twocolumn,pra,superscriptaddress]{revtex4-1}

\usepackage[dvips]{graphicx} 
\usepackage{enumerate}
\usepackage{epsfig}
\usepackage{subfigure}
\usepackage{xcolor}
\usepackage[T1]{fontenc}
\usepackage[expert]{mathdesign}
\usepackage{fullpage}
\usepackage{amsthm,amsfonts,amssymb,amscd,mathrsfs,eufrak,xspace,framed}
\usepackage{amsmath}
\usepackage{color}
\usepackage{setspace}
\usepackage{url}
\usepackage{enumitem}

\newcommand{\bra}[1]{\mbox{$\left\langle #1 \right|$}}
\newcommand{\ket}[1]{\mbox{$\left| #1 \right\rangle$}}

\begin{document}

\title{Trustworthiness of measurement devices in round-robin differential-phase-shift quantum key distribution}
\author{Zhu Cao}
\address{Center for Quantum Information, Institute for Interdisciplinary Information Sciences, Tsinghua University, Beijing, China}
\author{Zhen-Qiang Yin}
\email{yinzq@ustc.edu.cn}
\address{Key Laboratory of Quantum Information, University of Science and Technology of China, Hefei 230026, China\\
and Synergetic Innovation Center of Quantum Information $\&$ Quantum Physics, University of Science and Technology of China,\\
Hefei, Anhui 230026, China}
\author{Zheng-Fu Han}
\affiliation{Key Laboratory of Quantum Information, University of Science and Technology of China, Hefei 230026, China\\
and Synergetic Innovation Center of Quantum Information $\&$ Quantum Physics, University of Science and Technology of China,\\
Hefei, Anhui 230026, China}

\begin{abstract}
Round-robin differential-phase-shift quantum key distribution (RRDPS QKD) has been proposed to raise the noise tolerability of the channel. However, in practice, the measurement device in RRDPS QKD may be imperfect. Here, we show that, with these imperfections, the security of RRDPS may be damaged, by proposing two attacks for RRDPS systems with uncharacterized measurement devices. One is valid even for a system with unit total efficiency, while the other is valid even when a single photon state is sent. To prevent these attacks, either security arguments need to be fundamentally revised or further practical assumptions on the measurement device should be put.
\end{abstract}

\maketitle

{\it Introduction.} Quantum key distribution (QKD) enables an information-theoretic secure way of sharing a key between two distant parties Alice and Bob \cite{Bennett:BB84:1984,Ekert:QKD:1991}. In QKD, Alice sends quantum signals to Bob who measures to obtain raw keys. Due to disturbance in the channel or eavesdropping, the raw keys might not be identical or secure. So Alice and Bob perform error correction and privacy amplification to recover secure identical keys. In traditional QKDs, both error correction and privacy amplification costs are estimated from the error rates in the channel, which leads to an upper bound on the allowable error rate. For example, the error rate threshold for the celebrated BB84 QKD protocol is $25\%$ \cite{Bennett:BB84:1984}.

Recently, a QKD protocol, called round-robin differential-phase-shift (RRDPS) QKD \cite{sasaki2014practical} is proposed. This scheme and its passive equivalent scheme \cite{PRL114.180502}, were proven to be secure even under highly noisy channels, essentially removing the bound on the error rate. This desired property is achieved as they manage to directly compute the privacy amplification cost characterized by the phase error $e_{ph}$ without estimating the disturbance in the channel.

However, practical implementations of the RRDPS schemes \cite{PRL114.180502,RRDPSexp1,RRDPSexp2,RRDPSexp3} may be imperfect. For example, in the security proof of the RRDPS schemes \cite{sasaki2014practical,PRL114.180502}, Bob's measurement device is assumed to be a well-defined projective measurement acting on incoming single photon states. However, this assumption may be spoiled in practice because current detectors can malfunction under bright pulse illumination \cite{RRDPSexp1}. As another example, the beam splitter in the measurement device is assumed to have a transmission-reflection ratio of one half. However, in practice the transmission-reflection ratio of the beam splitter varies with different wavelengths and can thus be manipulated by adjusting the wavelength \cite{PhysRevA.84.062308}.

In this work, we show that these imperfections in the measurement device may call the security of RRDPS into question.
This is not obvious because the key security argument of RRDPS protocol, is that Alice's key is completely determined by the density matrix of her prepared state and Bob's $r$ is chosen by himself instead of the measurement device. These are not changed when the measurement device is imperfect. To model the imperfections, we consider the worst case scenario that Bob's measurement device is untrusted, and then propose two attacks to break the security.  One must impose that Bob's untrusted measurement device cannot communicate the measured keys with the adversary Eve outside Bob's environment, because otherwise no secure keys can be established. This can be done by, e.g., putting a shield around Bob's workstation.

In the following, we will first describe the untrusted measurement scenario in detail, and then describe our two attacks. Finally, we conclude with a few interesting open questions.

\begin{figure}[hbt]
\centering
\includegraphics[width=7cm]{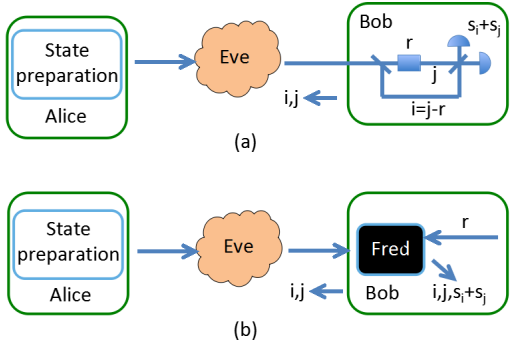}
\caption{(a) Original RRDPS \cite{sasaki2014practical}. Alice sends a pulse train to Bob, who splits the pulse train into two paths, adds a delay $r$ to
one of the paths, measures the interference to obtain $s_i\oplus s_j$, $i$, $j$, and publicly announces $i$, $j$.
(b) Untrusted measurement variant. Instead of performing the operations himself, Bob gives an untrusted measurement device, Fred, the pulse train along with a random $r$, and Fred outputs $s_i\oplus s_j$, $i$, $j=i+r$. One possible implementation of Fred could be exactly the same as the corresponding part of the original RRDPS.}
\label{fig:illus}
\end{figure}

{\it Untrusted measurement RRDPS.}
In the original RRDPS protocol, as shown in Fig.~\ref{fig:illus}(a), Alice's preparation of random bits sequence $s_1,\cdots, s_L$ and corresponding $L$-pulse state can be regarded as the following
operations equivalently. Alice firstly prepares the quantum state
\begin{equation}
\ket{\Psi}_1= \frac{1}{2^{L/2}} \prod\limits_{k=1}^L ( \ket{0}_k + (-1)^{\hat{n}_k}\ket{1}_k  )\ket{\Psi}_B,
\end{equation}
where $\ket{}_k$ represents Alice's ancillary qubits, $\hat{n}_k$ is the photon number operator acting on $\ket{\Psi}_B$, and $\ket{\Psi}_B$ is the encoding state which will be sent to Bob via a quantum channel. Alice's bit $s_k$ can be viewed as the output of $Z$ basis measurement $\{\ket{0}\bra{0},\ket{1}\bra{1}\}$ on ancilla $\ket{k}$.
Next, Alice sends the quantum state to Bob. Bob splits the input to two paths, applies a random delay $r$ to one of the paths, obtains $i$, $j=i+r$, $s_i \oplus s_j$ through detectors' outcomes and announces $i$ and $j$ to Alice. On receiving $i$ and $j$, Alice performs a controlled-NOT gate to her ancilla $\ket{}_i$ and $\ket{}_j$, in which $\ket{}_i$ is the control qubit while $\ket{}_j$ is the target. Then Alice can measure $\ket{}_j$ with the $Z$ basis to obtain her sifted key bit $s_i\oplus s_j$.

When the measurement is untrusted, as shown in Fig.~\ref{fig:illus}(b), Alice prepares the same state and sends it to Bob. After receiving the incoming state, Bob inputs a random delay value $r$ to his untrusted device, obtains $i, j$ $(j-i = r)$, and an estimated value of $s_i\oplus s_j$. The rest procedures are the same, where Bob publicly announces $i,j$ and Alice performs the controlled-NOT gate accordingly. Here untrusted means that Bob's device is an unknown quantum measurement device, called Fred, which could be manufactured by the adversary Eve. We assume here that Fred cannot actively disclose any information outside of Bob's environment bypassing Bob (e.g. by sending light signals). This assumption in particular eliminates the possibility of Fred telling Eve, who is outside, the measurement result $s_i\oplus s_j$ because Bob never announces it.

{\it Proposed attack 1.}
Before describing the first attack, we settle down some useful notations. For a function $f(n)$, $\Theta(f(n))$ stands for a function $g(n)$ which satisfies $c_1f(n)\le g(n)\le c_2f(n)$ for some constants $c_1,c_2$ independent of $n$, while $\omega(f(n))$ stands for a function $g(n)$ which satisfies $g(n)>cf(n)$ for any constant $c$.

The key idea of the attack is that Eve and Fred can cooperate to announce the $(i,j)$ pair. This is fundamentally different from the original RRDPS. To see the difference, examine the following example. Suppose the range of $i,j$ is $\{1,2,3,4\}$. In the original RRDPS, Eve chooses $i$. Then Bob chooses a random $j\not =i$ and thus has three choices of $j$. However, when Bob's measurement device is held by another adversary Fred who cooperates with Eve, the adversaries could restrict the $(i,j)$ pair output to $(1,2),(1,4),(2,4)$. For any $i$ that can be announced, $j$ only has two choices and thus not uniformly random anymore.

The setting of the attack is as follows. The pulse train contains $n$ pulses and Alice sends a pulse train containing $(n-1)/2$ photons. Note that according to the state preparation of RRDPS, each photon is in the state
\begin{equation}
\ket{\psi}= \frac{1}{\sqrt n}\sum\limits_{i=1}^n (-1)^{s_i}\ket{i}
\end{equation}
Eve first splits these $(n-1)/2$ photons and performs a passive interference with each photon. Then Eve obtains the relative phases of $(n-1)/2$ random pairs of $(i,j)$, with details in Appendix \ref{app:passive}.

A critical property of the relative phase is its transitivity. This can be easily shown as follows. If  both $s_i\oplus s_j$ and $s_j \oplus s_k$ are known, then
\begin{equation}
s_i \oplus s_k= (s_i\oplus s_j)\oplus (s_j \oplus s_k)
\end{equation}
is also known. For ease of presentation, we construct a graph with $n$ nodes corresponding to the $n$ pulses in the pulse train. If Eve detects the relative phase of $(i,j)$, add an edge between node $i$ and $j$. Then by the aforementioned property, if node $i$ and node $k$ can be connected by a path of nodes, their relative phase is known. Furthermore, by a well-known random graph result \cite{erdos}, since each edge appears with probability
\begin{equation}
p=\frac{\frac{n-1}{2}}{\binom{n}{2}}=\frac{1}{n}.
\end{equation}
there exists a connected component of size $\Theta (n^{2/3})$. Thus the relative phases between the $\Theta (n^{2/3})$ nodes in this connected component are all known by Eve. For a moderate size of $n=1000$, simulations show that with $p=1/40$ edge probability, there exists a connected component of size at least 200 at a $99\%$ confidence level.

Eve proceeds to send the corresponding indices of these $m=\Theta (n^{2/3})$ nodes and their phases to Fred. Denote the indices as $a_1,\dots, a_m$. By calculations of the probability in Appendix \ref{app:prob}, we show that with probability approaching 1, $\{|a_i-a_j|, 1\le i,j\le n\}$ will contain almost all numbers in the set $\{1,\cdots,n-1\}$.
For a moderate size $m=200$ and $n=1000$, simulations show that at a $99\%$ confidence level, at least $95\%$ of elements in $\{1,\cdots,n-1\}$ appear as differences of $a_i$.
Thus when Bob inputs a delay $r$ to Fred, he could announce a corresponding pair $a_i,a_j$ such that
\begin{equation}
|a_i-a_j|=r,
\end{equation}
and also outputs $s_{a_i}\oplus s_{a_j}$ to Bob. However, since $a_i,a_j$ are publicly announced, Eve, who is outside Bob's environment, can also recover $s_{a_i}\oplus s_{a_j}$ perfectly without causing any disturbance.

Finally, we note that, if some delay $r$ is absent, Fred could announce loss when Bob inputs this $r$. So in the asymptotic limit, the existence of a majority of delays from $\{1,\cdots,n-1\}$ could already guarantee Fred to output with almost unit probability.

Now we turn to the implications of this attack. Let us examine the phase error formula for RRDPS protocol. In the original work \cite{sasaki2014practical}, the phase error formula is $e_{ph}=n_{ph}/(L-1)$, where $L$ is the number of pulses in the pulse train and $n_{ph}$ is the number of photons that Alice sends. We plug in $n_{ph}=(n-1)/2$ and $L=n$ and get $e_{ph}=1/2$. This means that there are no secure keys and does not contradict with the fact that Eve has obtained all the information that Alice and Bob share.

However, the phase error in the original paper is not tight. Indeed, the phase error is later improved \cite{PRL114.180502, arXiv1505} and is given by $e_{ph}=[1-(1-2/L)^{n_{ph}}]/2$. The improvement relies on the fact that when Eve chooses $i$ and Bob chooses a random $j\not =i$ for multiple photons, an phase error happens only with an odd number of photons. By plugging in $n_{ph}=(n-1)/2$ and $L=n$, we have $e_{ph}=(1-1/e)/2$ when $n$ is large. This implies that secure keys can be generated and a contradiction occurs. This shows that in the untrusted measurement setting, with even unit efficiency, careful study of the phase error is needed since the model is no longer equivalent to the original RRDPS protocol. Through the proposed attack, Eve can learn key bits without introducing any additional error rates. In the following, we propose another attack, in which Eve can learn key bits with the help of additional error rates when total loss is larger than 3dB.

{\it Proposed attack 2.} In the second attack, Eve and Fred will obtain Alice's key bit with certainty, but introduce $50\%$ bit error rate and $50\%$ channel loss between Alice and Bob.

Consider Alice prepares the following single photon encoding state
\begin{equation}
\label{Psi_1}
\begin{split}
\ket{\Psi}_1=\frac{1}{\sqrt{L}}&(e^{i\alpha_1}\ket{1}_1\ket{0}_2\ket{0}_3\cdots\ket{0}_L\\
&+e^{i\alpha_2}\ket{0}_1\ket{1}_2\ket{0}_3\cdots\ket{0}_L\\
&+e^{i\alpha_3}\ket{0}_1\ket{0}_2\ket{1}_3\cdots\ket{0}_L\\
&+...\\
&+e^{i\alpha_L}\ket{0}_1\ket{0}_2\ket{0}_3\cdots\ket{1}_L).
\end{split}
\end{equation}
Then Eve launches an attack in the following way:

\begin{equation}
\label{Psi_2}
\begin{split}
\ket{\Psi}_2=\frac{1}{\sqrt{L}}&(e^{i\alpha_1}\ket{1}_1\ket{0}_2\ket{0}_3\cdots\ket{0}_L\ket{1}_E\\
&+e^{i\alpha_2}\ket{0}_1\ket{1}_2\ket{0}_3\cdots\ket{0}_L\ket{2}_E\\
&+e^{i\alpha_3}\ket{0}_1\ket{0}_2\ket{1}_3\cdots\ket{0}_L\ket{3}_E\\
&+...\\
&+e^{i\alpha_L}\ket{0}_1\ket{0}_2\ket{0}_3\cdots\ket{1}_L\ket{L}_E),
\end{split}
\end{equation}
where, $\ket{n}_E$ is the ancilla of Eve and $_E\bra{n}m\rangle_E=\delta_{mn}$. Eve sends this ancilla $E$ to Bob's device (Fred) and retains Alice's encoding state $|\rangle_1|\rangle_2...|\rangle_L$ in her quantum memory.
Fred simply measures $E$ according to Bob's input $r$. For example, if $r=1$, Fred measures $E$ with the basis $(\ket{1}_E\pm \ket{2}_E, \ket{2}_E\pm \ket{3}_E,...)$, which is the same as the $r$-delay interference and will output a pair of $i$ and $j$ definitely. If Fred obtains $\ket{i}_E+ \ket{j}_E$, Fred will inform Bob the $i$, $j$, and a random key bit. If Fred obtains $\ket{i}_E- \ket{j}_E$, Fred will inform Bob that there is no click in this run. For instance, Fred observes $(\ket{1}_E+ \ket{2}_E)$ in this run, then Alice's encoding state becomes

\begin{equation}
\label{Psi_3}
\begin{split}
\ket{\Psi}_3=\frac{1}{\sqrt{2}}&(e^{i\alpha_1}\ket{1}_1\ket{0}_2\ket{0}_3\ket{0}_L\\
&+e^{i\alpha_2}\ket{0}_1\ket{1}_2\ket{0}_3\ket{0}_L).
\end{split}
\end{equation}
Since this state is retained by Eve, she can learn Alice's sifted key bit $\alpha_1\oplus \alpha_2$ easily. Though this attack introduces high error rate, it implies that the security of RRDPS in the untrusted measurement scenario will depend on the error rate, which is completely different from original RRDPS protocol. Hence, one may conjecture that Eve's information on key bits depends on the error rate in the untrusted measurement setting, while the original RRDPS protocol features that monitoring signal disturbance is not necessary for placing a bound on Eve's information.

The error rate $50\%$ can be reduced by mixing performing and not performing this attack together with probability $1/2$ each. Then the error rate drops to $25\%$, while the same feature, channel-dependent phase error, still holds.
Hence, when Bob's device is untrusted, the RRDPS protocol is not secure and monitoring signal disturbance would be necessary. This finding may shed lights on the future security proof of RRDPS in untrusted measurement scenario.

{\it Discussion.} In summary, we have shown that the security of RRDPS protocol can be damaged when the measurement device is not perfect and consequently does not satisfy the assumptions of the protocol. We prove this by giving two concrete attacks for RRDPS with completely untrusted measurement device. In addition to showing a warning sign for the trustworthiness of the measurement device, our work raises two interesting questions:

The first question is how to prove the security of RRDPS with untrusted measurement.
Our result shows that in the untrusted measurement setting, the security analysis is significantly different from the original RRDPS, and thus calls for more scrutiny in the analysis of the untrusted measurement setting. We show in Appendix \ref{app:security} that it can be proven to be secure in the perfect case where there is no error and no loss.
 We therefore conjecture that it could also be made secure under a general lossy and erroneous channel when one correctly characterizes the information leakage. But the security proof may require transforming the experiment setup to a completely different intermediate model.

The second question is motivated by the fact that our untrusted measurement assumption may not be satisfied by a current-technology-hackable measurement device. Hence, it is interesting to see how our attacks can be modified and implemented on a current experimental RRDPS setup.

The authors would like to thank X. Ma for enlightening discussions. This work was supported by the National Basic Research Program of China (Grants No. 2011CBA00200 and No. 2011CB921200), National Natural Science Foundation of China (Grants No. 61475148) and the 1000 Youth Fellowship program in China.

\appendix
\section{Passive interference}
\label{app:passive}
In the passive interference, Eve prepares a superposed state with plain phases,
\begin{equation}
\ket{\psi_0}=\frac{1}{\sqrt n}\sum\limits_{i=1}^n \ket{i},
\end{equation}
and interferes with Alice's photon with a beam splitter which results in a mixture of
the two click events \cite{PRL114.180502},
\begin{equation}
[1+(-1)^{s_i+s_j}] c_i c_j,\quad [1-(-1)^{s_i+s_j}] c_i d_j
\end{equation}
where $c_i$ denotes the first detector clicks at the $i$-th time and $d_j$ denotes the second detector clicks at the $j$-th time. Thus by examining whether the two clicks are from the same detector or from different detectors, Eve can determine the value of $s_i\oplus s_j$ for a random pair $(i,j)$.

\section{Equidistribution of differences}
\label{app:prob}
In this section, we will prove a technical result which is involved in the attack analysis,
namely
\begin{equation}
\lim_{n\to \infty} \mathbb{E}\left[ \frac{\#\{|a_i-a_j|,1\le i,j\le m\}}{n}\right]=1,
\end{equation}
where $m=\Theta (n^{2/3})$ and $a_1,\cdots, a_m$ are chosen at random from $\{1,\dots, n\}$.
Actually we will prove a stronger result which replaces $m=\Theta (n^{2/3})$ by $m=\omega (\sqrt{n})$. It is stronger because if the conclusion holds for a specific $m$, it naturally also holds for bigger $m$.
The conclusion is intuitively true because $\omega (\sqrt{n})$ numbers have more than
\begin{equation}
\binom{2\sqrt{n}}{2}>n
\end{equation}
differences and these differences have sufficient randomness in $\{0,\dots, n-1\}$.

Now let us start the proof. First fix a constant $\epsilon>0$. Let $A=\{a_i:i<m/2\} \cap \{1,\cdots, \epsilon n\}$. In expectation, $A$ contains $\epsilon m/2=\omega(n)$ elements. By concentration, we have $ |A| \ge \sqrt{n}$ with high probability.

Now suppose that $|A| \ge \sqrt{n}$ holds, i.e., there are $\sqrt{n}$ distinct values in  $\{1,\cdots, \epsilon n\}$ for the first $m/2$ elements. Then for the last $m/2$ elements, they are independent and each element has probability $\sqrt{n}/n$ to be distance $d$ away from some element in $A$ for any $d\in [1,n-\epsilon n]$. Thus the probability that none of the last $m/2$ elements are distance $d$ from some element in $A$ is $ (1-1/\sqrt{n})^{m/2}$, which vanishes as $n$ goes to infinity.

So, using linearity of expectation for all $d\in [1,n-\epsilon n]$, we have
\begin{equation}
\lim_{n\to \infty} \mathbb{E}\left[ \frac{\#\{|a_i-a_j|,1\le i,j\le m\}}{n}\right]\ge 1-\epsilon,
\end{equation}
for any $\epsilon >0$. Since $\epsilon$ can be arbitrarily small, our result holds and the proof is complete.

{\it Remark.} Note that $m=\omega(\sqrt{n})$ is actually tight. One cannot hope to improve the bound to $m=\Omega(\sqrt{n})$. The reason is as follows. Suppose there is some constant $c$ such that $m=c \sqrt{n}$, then in $[1,n/c^3]$ and $[n-n/c^3,n]$, about $2\sqrt{n}/c^2$ numbers are selected. Then differences in $[n-n/c^3,n]$ will appear for about $n/c^4$ times. Thus when $n\to \infty$, the upper bound on the expectation value of the number of differences dividing $n$ is $1- 1/c^3+1/c^4$, not 1.

\section{Security for the perfect case}
\label{app:security}

For simplicity, we consider the case that $L=3$ in this proof, since this proof can cover larger $L$ cases easily. And only collective attack is considered here. Alice's encoding state can be written as $\ket{\Psi}=(-1)^{k_1}\ket{1}+(-1)^{k_2}\ket{2}+(-1)^{k_3}\ket{3}$.  Besides no loss and error, we also assume that for the events that Fred outputs $i$, $j$, Alice finds that $k_l(l\neq i,j)$ have equal 0s and 1s conditioned on $(k_i,k_j)=(a,b)$ for all $a,b\in\{0,1\}$.

The general collective attack model is that Eve applies a unitary transformation $U_E$ to the traveling photon and her ancilla $\ket{E}$. Note that after Eve's operation $U_E$, the incoming photon will be transformed to an unknown and arbitrary information carrier, labeled as $\ket{F}$, sent to Fred. Then depends on Bob's input $r$, Fred uses unitary transformation $U^{(r)}_F$ acting $\ket{F}$ to generate $\ket{i}+(-1)^{{k_i}\oplus k_i}\ket{j}$ to Bob. When Fred generates $\ket{1}\pm\ket{2}$, the density matrix of $\ket{E}$ is given by
\begin{equation}
\begin{aligned}
\label{attack1}
\rho_E&=\sum_{k_3}tr[
(\ket{1}\bra{1}+\ket{2}\bra{2})\\
&P\{U_E((-1)^{k_1}\ket{1}+(-1)^{k_2}\ket{2}+(-1)^{k_3}\ket{3})\\
&\ket{E}\ket{F}\}]\\
&=tr[
(\ket{1}\bra{1}+\ket{2}\bra{2})\\
&P\{U_E((-1)^{k_1}\ket{1}+(-1)^{k_2}\ket{2})\ket{E}\ket{F}\}\\
&+P\{U_E\ket{3}\ket{E}\ket{F}\}],
\end{aligned}
\end{equation}
where, $P\{\ket{x}\}=\ket{x}\bra{x}$. From \eqref{attack1}, to analyze Eve's ancilla $\ket{E}$, Alice's encoding state is equivalent to a mixture of components $P\{(-1)^{k_1}\ket{1}+(-1)^{k_2}\ket{2}\}$ and $P\{\ket{3}\}$. Evidently, in perfect case, the events that Fred outputs $1$ and $2$ can only correspond to that Alice emits $(-1)^{k_1}\ket{1}+(-1)^{k_2}\ket{2}$. Furthermore, in perfect case, Fred's operation $U^{(1)}_F$ generates $(-1)^{k_1}\ket{1}+(-1)^{k_2}\ket{2}$ without error, which means
\begin{equation}
\begin{aligned}
\label{eve1}
&U_E( \ket{1}+ \ket{2} )\ket{E}=\ket{F_{12+}}\ket{E_{12+}},\\
&U_E( \ket{1}- \ket{2} )\ket{E}=\ket{F_{12-}}\ket{E_{12-}},
\end{aligned}
\end{equation}
and
\begin{equation}
\begin{aligned}
\label{fred1}
&U^{(1)}_F\ket{F_{12+}}=\ket{1}+ \ket{2},\\
&U^{(1)}_F\ket{F_{12-}}=\ket{1}- \ket{2}.
\end{aligned}
\end{equation}
On the other hand, when $r=2$, and Fred outputs $1$ and $3$, we have
\begin{equation}
\begin{aligned}
\label{eve2}
&U_E( \ket{1}+ \ket{3} )\ket{E}=\ket{F_{13+}}\ket{E_{13+}},\\
&U_E( \ket{1}- \ket{3} )\ket{E}=\ket{F_{13-}}\ket{E_{13-}},
\end{aligned}
\end{equation}
and
\begin{equation}
\begin{aligned}
\label{fred2}
&U^{(2)}_F\ket{F_{13+}}=\ket{1}+ \ket{3},\\
&U^{(2)}_F\ket{F_{13-}}=\ket{1}- \ket{3}.
\end{aligned}
\end{equation}
Since unitary transformation $U_E$, $U^{(1)}_F$, and $U^{(2)}_F$ must preserve inner product, it's easy to find the
module of the product of $ \ket{E_{13+}}$ and  $\ket{E_{13-}}$ must be 1, which implies that  $ \ket{E_{13+}}$ and  $\ket{E_{13-}}$ can only differ by a phase. Hence, Eve can not obtain any information on $k_1\oplus k_3$ through her ancilla $\ket{E}$. Similarly, we can prove that Eve cannot learn $k_1\oplus k_2$ and $k_2\oplus k_3$.

\bibliographystyle{apsrev4-1}

\bibliography{Bibliattack}

\end{document}